\documentclass{aa}

\usepackage{graphicx}
\usepackage{epsfig}
\usepackage{graphics}
\usepackage{subfigure}
\usepackage{amsmath}
\usepackage{enumerate}
\usepackage{natbib}

\begin{document}

\title{Particle transport in magnetized media around black holes and associated radiation}

\author{F. L. Vieyro\inst{1,2,}\thanks{Fellow of CONICET, Argentina} \and G. E. Romero\inst{1,2,}\thanks{Member of CONICET, Argentina}}
  
\institute{Instituto Argentino de Radioastronom\'{\i}a (IAR, CCT La Plata, CONICET), C.C.5, (1984) Villa Elisa, Buenos Aires, Argentina \and Facultad de Ciencias Astron\'omicas y Geof\'{\i}sicas, Universidad Nacional de La Plata, Paseo del Bosque s/n, 1900, La Plata, Argentina}

\offprints{F. L. Vieyro \\ \email{fvieyro@iar-conicet.gov.ar}}

\titlerunning{Particle transport around black holes}

\authorrunning{Vieyro \& Romero}

\abstract
{Galactic black hole coronae are composed of a hot, magnetized plasma. The spectral energy distribution produced in this component of X-ray binaries can be strongly affected by different interactions between locally injected relativistic particles and the matter, radiation and magnetic fields in the source.}
{We study the non-thermal processes driven by the injection of relativistic particles into a strongly magnetized corona around an accreting black hole.}
{We compute in a self-consistent way the effects of relativistic bremsstrahlung, inverse Compton scattering, synchrotron radiation, and the pair-production/annihilation of leptons, as well as hadronic interactions. Our goal is to determine the non-thermal broadband radiative output of the corona. The set of coupled kinetic equations for electrons, positrons, protons, and photons are solved and the resulting particle distributions are computed self-consistently. The spectral energy distributions of transient events in X-ray binaries are calculated, as well as the neutrino production.}
{We show that the application to Cygnus X-1 of our model of non-thermal emission from a magnetized corona yields a good fit to the observational data. Finally, we show that the accumulated signal produced by neutrino bursts in black hole coronae might be detectable for sources within a few kpc on timescales of years.}
{Our work leads to predictions for non-thermal photon and neutrino emission generated around accreting black holes, that can be tested by the new generation of very high energy gamma-ray and neutrino instruments.}
 
 
\maketitle

\section{Introduction}

Some physical processes that take place near a black hole can be inferred from X-ray observations of accreting binaries. There is strong evidence that the X-ray emission in some of these systems is powered by accretion onto a black hole. Accretion processes around compact objects are modeled by hydrodynamic equations of viscous differentially-rotating fluids. The \textsl{standard disk model} is the most famous model based on this set of equations, and was developed by \citet{shakura} and \citet{novikov}. The model predicts that the accreting gas forms a geometrically thin and optically thick disk, producing a quasi-blackbody spectrum due to thermal emission. The effective temperature of the accreting gas is in the range $10^5-10^7$ K, depending on the mass of the compact object and the accretion rate.

The observed X-ray spectrum, however, is too hard in many cases to have been produced by a standard disk alone, since the hard X-ray component corresponds to a temperature of $\sim 10^9$ K. 

To explain the complete X-ray spectrum of Cygnus X-1 --which is arguably the most well-studied black hole candidate in the Galaxy-- an extra component is usually added to the disk, the so-called \textsl{corona} (e.g., \citealt{dove1997,gier,poutanen1998}). In this context, the corona is coupled to the disk by the magnetic field, and the plasma is heated by reconnecting magnetic loops emerging from the disk \citep{galeev1979}. 

As a result, the soft photons emitted by the disk gain energy by successive Compton upscatterings in the corona. This process is known as Comptonization, and it is the most accepted mechanism to explain the broad-band hard X-ray spectra of Galactic black holes. Another feature that can be explained by the presence of the corona, is a hardening of the spectra at $\sim 10$ keV, which is attributed to the Compton reflection of hard radiation (emitted by the corona) from a cold material, i.e. the disk \citep{white1988,george1991}.

Since their early applications to Cygnus X-1 \citep{shapiro,ichimaru}, different models have attempted to explain the complete X-ray spectrum of black holes in binary systems. The observational data of the spectra are used to set some constraints on the geometry of the source. For example, the covering fraction of the hot cloud as viewed from the soft photon source can be estimated from the observed spectral slopes, and the reflection bump limits the solid angle subtended by the disk around the corona \citep{poutanen1997}. There are two specific geometries that seem to reproduce well most of the observed spectra: the \textsl{disk + corona `sombrero'} model and the \textsl{advection dominated accretion flow} model (ADAF).

In the `sombrero' model, the corona is added to the disk. It is usually assumed that the corona can be represented by a homogeneous spherical cloud of radius $R_{\rm{c}}$ around the compact object; the cold disk (with an inner radius $r_{\rm{in}}$) is truncated at a certain distance from the compact object in the hard state of Galactic black holes (\citealt{dove1997,poutanen1998}). The ratio $r_{\rm{in}}/R_{\rm{c}}$ depends on the amount of Compton reflection observed, and can be obtained from the energy balance and the electron/positron pair balance (\citealt{dove1997,poutanen1997}). In the case of Cygnus X-1, this ratio is estimated to be $r_{\rm{in}}/R_{\rm{c}} = 0.8 - 0.9$, which means that only a short fraction of the disk is within the hot cloud \citep{poutanen1997}. 

On the other hand, the ADAF model is a self-consistent solution of the hydrodynamic equations of viscous rotating flows (see, e.g.,  \citealt{ichimaru,narayan1994,narayan1995a,narayan1995b,abramowicz1995}). The main characteristic of this solution is that most of the energy is accreted onto the compact object and the gas is unable to cool efficiently, mainly because of the low density. Since the plasma is a poor radiator, the viscous energy is stored in the gas as thermal energy and the gas temperature becomes high. This causes the accreting gas to swell, hence the geometry of ADAFs is quasi-spherical. This geometry is similar to the geometry of the corona+disk `sombrero' model already mentioned above.

Besides the X-ray spectra observed in X-ray binaries (XRBs), sources such as Cygnus X-1 produce steady emission up to a few MeV \citep{McConnell,cadolle2006}, that is indicative of a non-thermal contribution to the spectral energy distribution (SED). \citet{li1996} treated simultaneously the transport and acceleration mechanisms for leptons obtaining good fits to the spectra of Cygnus X-1 observed by COMPTEL. In addition, \citet{li1997} demonstrated that under the physical conditions expected in the surroundings of accreting black holes, it is possible to accelerate electrons out of a Maxwellian distribution, resulting in a non-thermal tail. 

More recently, several works have been devoted to studying the effects of non-thermal populations of electrons in Galactic black hole coronae (e.g., \citealt{belmont2008,malzac2009,vurm2009}), as well as in accreting supermassive black holes (e.g., \citealt{belmont2008,veledina2011}). These works, however, have not studied the effects of a hadronic component in the source. 

In \citet{flor01}, it was shown that the presence of hadrons can also explain the non-thermal emission detected in Galactic black holes. In the present work, we show the results of a theoretical study of the effects of the injection of non-thermal particles, both electrons and protons, in a magnetized corona around a black hole, that has many refinements compared to previous works. In particular, a self-consistent treatment of photon and particle transport is now presented. We solve the set of coupled kinetic equations for all types of particles, including photons, hence the treatment of absorption is straightforward. We attempt to estimate the SED produced by these relativistic particles, and explain the origin of the non-thermal tail observed in some XRBs \citep{McConnell,cadolle2006,jourdain2011}. 

The completion of IceCube opens new possibilities for neutrino detection, so we also study the neutrino production in the source.

The structure of this paper is as follows. In the next section, we present the basic model used to obtain the values of the relevant parameters, and fix the characteristics of the medium where the non-thermal particles are injected. In Sec. \ref{treatment}, we discuss the treatment of the physical processes, along with the details of the numerical method used to solve the corresponding equations. We first study the system in a steady state, to test our code and reproduce the results obtained in \citet{flor01}.  In Sec. \ref{flare}, we then consider the case of a transient event, including the time-dependence of the transport equations. Finally, we estimate the electromagnetic emission as well as the neutrino flux.

\section{Basic model}

In accordance with estimates for Cygnus X-1, we assume a black hole of mass $M_{\rm{BH}}=14.8 M_{\odot}$ \citep{orosz2011}. We study the system in the low-hard state, which is the state where the X-ray spectrum is dominated by the coronal emission. 

\begin{figure}
\centering
\includegraphics[clip,width=0.49\textwidth, keepaspectratio]{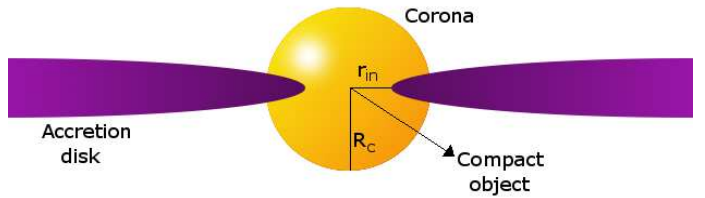}
\caption{Schematic geometry of the system in the low-hard state.}
\label{fig:lowHard}
\end{figure}

The size of the region where the hard radiation is produced, is limited by the variability observed in the spectra of Galactic black holes. The minimum variability timescale is on the order of milliseconds, hence in the hard state the corona lies within $\sim 20-50$ $r_{\rm{g}}$ \citep{poutanen1998}, where $r_{\rm{g}}$ is the gravitational radius ($r_{\rm{g}}=GM/c^2$). We considered a spherical corona with a size of $R_{\rm{c}} = 35r_{\rm{g}}$ (see Fig. \ref{fig:lowHard}), and assumed that the luminosity of the corona is 1 \% of the Eddington luminosity \citep{esin1997}, which results in $L_{\rm{c}} = 1.9 \times 10^{37}$ erg s$^{-1}$. 

In ADAF models, ions and electrons interact only through Coulomb collisions and there is no coupling between the two species. In this case, the plasma has two temperatures, with the ion temperature ($T_{i}=10^{12}$ K) being much higher than the electron temperature ($T_{e}=10^{9}$ K) (e.g., \citealt{narayan1995a,narayan1995b}).

We initially consider that the system is in steady state, thus the equipartition of energy between the different components of the system is a reasonable assumption \citep{esin1997}. The power of the jet observed in the low-hard state of X-ray binaries is related to the magnetic field, which in turn is coupled to the corona. We then assume equipartition between the magnetic energy density and the bolometric photon density of the corona $L_{\rm{c}}$ (as done by, e.g., \citealt{bednarek2007}). At the same time, since the magnetic field launches the plasma into a jet, we also assume equipartition between the magnetic and kinetic energy densities (see e.g., \citealt{zdziarski1998}). These conditions allow us to obtain the values of the main parameters. Specifically, the conditions are
	
	\begin{equation}
		\frac{B^2}{8\pi}=\frac{L_{\rm{c}}}{4\pi R_{\rm{c}}^2c} ,
	\end{equation}
			
			\begin{equation}
				\frac{B^2}{8\pi}=\frac{3}{2}n_{e}kT_{e}+\frac{3}{2}n_{i}kT_{i} ,
			\end{equation}
			
\noindent where $B$ is the value of the mean magnetic field in the corona and $n_{e}$, $n_{i}$ are the electron and ion plasma densities, respectively.

The X-ray emission of the corona is characterized by a power law in photon energy $\epsilon$ with an exponential cut-off at high energies (e.g. \citealt{romero2002})

			\begin{equation}
				n_{\rm{ph}}(\epsilon)=A_{\rm{ph}} \epsilon^{-\alpha}  e^{-\epsilon/\epsilon_{\rm{c}}} \textrm{erg$^{-1}$ cm$^{-3}$.}
			\end{equation}
			
\noindent We adopt $\alpha=1.6$ and $\epsilon_{\rm{c}}=150$ keV, which are typical of Cygnus X-1 (\citealt{poutanen1997}). The photon field of the accretion disk is modeled as a blackbody of temperature $kT_{\rm{d}}=0.1$ keV. Both the X-ray emission of the corona and the radiation field of the disk are considered as seed photon sources for Compton scattering and photomeson production in relativistic particle interactions.

In ADAF models, particles are advected onto the compact object at a mean velocity of $v \sim 0.1c$. The accretion timescale is about a second, which is considerably shorter than the cooling rate for hadrons \citep{flor01}. As a result, only high energy protons are able to lose a significant amount of energy, and most of the power injected in terms of relativistic hadrons falls into the black hole. Since most of the electron/positron pairs are produced by hadronic interactions, in advective-dominated coronae the high energy emission is likely produced in the jet \citep{markoff2001,vila2010,vila2012}.

In \citet{flor01}, a static corona model was shown to be capable of producing the non-thermal emission of Cygnus X-1. In this work, we consider only this possibility. In a static corona, which is supported by magnetic pressure, the relativistic particles can escape mainly by diffusion \citep{beloborodov1999,nayakshin1999}. In the Bohm regime, the diffusion coefficient is $D(E)=r_{\rm{g}}c/3$, where $r_{\rm{g}}=E/(eB)$ is the gyroradius of the particle. The diffusion rate is

	\begin{equation}\label{eq:diff}
				t_{\rm{diff}}^{-1}=\frac{2D(E)}{R_{\rm{c}}^2} .
		\end{equation}
		
We consider the injection of the non-thermal particle distributions of both electrons and protons in this corona. The mechanism of particle acceleration in black hole coronae is likely related to magnetic reconnection, which is essentially a topological reconfiguration of the magnetic field caused by a change in the connectivity of the field lines. Several works have been published on particle acceleration through magnetic reconnection (e.g. \citealt{schopper1998,zenitani2001,bette2010}) The basic idea is that a first-order Fermi mechanism takes place within the reconnection zone caused by two converging magnetic fluxes of opposite polarity that move toward each other with a velocity $v_{\rm{rec}}$ \citep{bette2005}. The resulting injection function of relativistic particles is a power-law with an index $\Gamma \sim 2.2$.  

The detailed analysis of \citet{drury2012} supports the idea that magnetic reconnection can lead to type I Fermi acceleration in a similar way as diffusive acceleration is driven by shocks. According to Drury, the spectral index does not have a universal value of $2.2$; it instead lies somewhere in the range $1 \leq \Gamma \leq 3$. Despite this caveat, and given the uncertainties in the acceleration models, we adopt the value of $\Gamma = 2.2$, which is consistent with all simulations implemented so far. 

In our model, the acceleration mechanism is not included as a term of the energy gain in the transport equation, but used to fix the injection function of primary electrons and protons and determine the maximum energies that relativistic particles can achieve. 

As for standard first order Fermi acceleration, the acceleration rate $t^{-1}_{\rm{acc}}=E^{-1}dE/dt$ for a particle of energy $E$ in a magnetic field $B$ is given by

\begin{equation}\label{eq:accrate}
	t^{-1}_{\rm{acc}}=\frac{\eta ecB}{E},
\end{equation}

\noindent where $\eta$ is a parameter that characterizes the efficiency of the mechanism in the magnetized plasma. It is given by \citep{drury1983,vilaaha,delValle2011}

\begin{equation}
\eta \sim 0.1 \frac{r_{\rm{g}}c}{D} \Big( \frac{v_{\rm{rec}}}{c} \Big)^2.
\end{equation}

\noindent The reconnection speed in violent reconnection events is $v_{\rm{rec}} \sim v_{\rm{A}}$ \citep{lazarian1999,kowal2011}, where $v_{\rm{A}}$ is the Alfv\'en velocity, given by

\begin{equation}
v_{\rm{A}} =\sqrt{ \frac{B^2}{4\pi m_{p} n} }.
\end{equation}

\noindent In the corona, the Alfv\'en speed is $\sim 0.5 c$, yielding an acceleration efficiency of $\eta \sim 10^{-2}$. 

The power available in the magnetized system can be estimated by \citep{delValle2011}

\begin{equation}
L = \frac{B^2}{8\pi } A v_{\rm{A}},
\end{equation}

\noindent where $A \sim 4\pi R_{\rm{c}}^2$. This yields a power available for non-thermal processes of $\sim 15$ \% $L_{\rm{c}}$. The total power injected into relativistic protons and electrons, $L_{\rm{rel}}$, is assumed to be a fraction of the luminosity of the corona, $L_{\rm{rel}} = q_{\rm{rel}} L_{\rm{c}}$, with $q_{\rm{rel}} < 0.15$. The way in which energy is divided between hadrons and leptons is unknown. To deal with this uncertainty, it is useful to define the parameter $a$ as the ratio of the power injected in protons to the one injected in electrons, $a=L_{p}/L_{e}$. Following \citet{flor01}, we consider models with $a = 100$ (proton-dominated case, as for Galactic cosmic rays) and $a=1$, that is, models with the same power injected in both hadrons and leptons. In our model, the injection function is assumed to be both homogeneous and isotropic.

We model the escape of radiation from the region using the treatment described in \citet{coppi1992}

\begin{equation}
t_{\rm{esc}}(E_{\gamma}) = \frac{R_{\rm{c}}}{c}\left[ 1+ \tau_{\rm{KN}} f(E_{\gamma}) \right] ,
\end{equation}

\noindent where

\begin{equation}
			f(E_{\gamma}) = 
			 \begin{cases}
				1 & \textrm{for } x \leq 0.1,\\
				\frac{1-E_{\gamma}/m_{e}c^2}{0.9}  &   \textrm{for } 0.1 < x < 1,\\
				0  &   \textrm{for } x \geq 1,\\ 
			\end{cases}
\end{equation}

\begin{equation}
\tau_{\rm{KN}} = 2R_{\rm{c}} <\sigma_{\rm{KN}} E_{e^{\pm}}> N_{e^{\pm}},
\end{equation}

\noindent $x=E_{\gamma}/m_{e}c^2$, $\sigma_{\rm{KN}}$ is the Klein-Nishina cross-section, $N_{e^{\pm}}$ is the number density of pairs defined by $N_{e^{\pm}} = \int dE_{e^{\pm}} N_{e^{\pm}}(E_{e^{\pm}})$, and $<\sigma_{\rm{KN}} E_{e^{\pm}}>$ represents an average over the particle distribution such that photons with energy $E_{\gamma} > m_{e}c^2$ escape out of the source in a time $R_{\rm{c}}/c$. Table \ref{table} summarizes the values of the relevant parameters in our model.

\begin{table}
    \caption[]{Main parameters of the model.}
   	\label{table}
   	\centering
\begin{tabular}{ll}
\hline\hline 
Parameter & Value\\ [0.01cm]
\hline   
$M_{\rm{BH}}$:  black hole mass [$M_{\odot}$]										& $14.8$\tablefootmark{a}	\\
$R_{\rm{c}}$:   corona radius [$r_{\rm{g}}$] 										& $35$ 	                	\\
$r_{\rm{in}}/R_{\rm{c}}$: inner disk/corona ratio               & $0.9$\tablefootmark{b} 	\\
$T_{e}$:        electron temperature [K] 						  					&	$10^9$								\\
$T_{i}$:        ion temperature [K] 														&	$10^{12}$ 						\\
$\epsilon_{\rm{c}}$:   X-ray spectrum cut-off [keV]							& $150$       					\\
$\alpha$: 			X-ray spectrum power-law index    							& $1.6$									\\
$\eta$: 				acceleration efficiency 												& $10^{-2}$							\\
$B_{\rm{c}}$: 	magnetic field [G]				 											& $5.7 \times 10^5$			\\
$n_{i},n{e}$:   plasma density [cm$^{-3}$] 											& $6.2 \times 10^{13}$	\\
$kT$:						disk characteristic temperature [keV] 	  			& $0.1$									\\

\hline  \\[0.005cm]
\end{tabular}	
\tablefoot{
\tablefoottext{a}{Value for Cygnus X-1 \citep{orosz2011}.}\\
\tablefoottext{b}{
This is the typical value in models where the main source of seed photons for Comptonization are provided by the disk \citep{poutanen1998,haardt1993}. In our model, there are other relevant sources of photons that can relax the condition $r_{\rm{in}}/R_{\rm{c}}=0.9$; however, we only use this parameter to constrain the size of the corona, which for Cygnus X-1 cannot differ significantly from the one adopted in our work \citep{poutanen1998}}.
 } 
\end{table}

\section{Treatment of radiative processes}\label{treatment}

Computing the SEDs of black hole coronae, as well as other magnetized plasmas, is a complex task, since it must include, among other issues, a detailed knowledge of the plasma characteristics and the microphysical processes. The first method used to treat this problem was a Monte Carlo simulation (e.g., \citealt{aharonian1985,stern1995,pilla1997}). The main problem for this approach is the small number of high-energy photons, which leads to low quality photon statistics. On the other hand, it is usually easy to model the radiative transfer processes (see \citealt{pellizza2010} for a three-dimensional code).

A second method for estimating the spectra of compact sources involves solving the kinetic equations (e.g., \citealt{lightman1987,coppi1990,coppi1992}). The different interactions of particles with the fields of the source ensure that it is unavoidable to study a wide range of particle energies, hence the computation of radiative processes is quite complicated using this approach. The main advantage of this method is that the transport of photons is solved self-consistently. Since, in general, these equations are solved numerically, the availability of computational resources over the past decade has allowed the significant improvement of this approach (see, e.g., \citealt{aharonian2003,malzac2009,poutanen2009,vurm2009}). 

To complete the treatment initiated in \citet{flor01}, we take the second approach. As mentioned in the previous section, we are interested in the study of the injection of non-thermal particle distributions of electrons and protons in the system. Once protons are injected into the corona, they interact with both the photon and matter fields, producing pions. In addition, the charged pions decay producing muons, so we also take into account the presence of these transient particles.

An accurate description of a hot, magnetized plasma such as the corona should also treat the processes of pair production and annihilation, hence we include electron/positron pairs. 

The main channel for secondary pair production in our model is photon-photon annihilation. The most important background photon field for pair creation is the thermal X-ray radiation of the corona. The energy spectrum of pairs has been studied, for example, by \citet{aharonian1983} and \citet{boettcher1997}. Under the conditions $\epsilon \ll m_{e}c^2 \leq E_{\gamma}$, the pair emissivity $Q_{\gamma\gamma\rightarrow e^{\pm}}(E_{e})$ (in units of erg $^{-1}$ s$^{-1}$ cm$^{-3}$) can be approximated by the expression

	\begin{align}
		Q_{\gamma\gamma\rightarrow e^{\pm}}(E_{e^{\pm}}) & =  \frac{3}{32} \frac{c\sigma_{\rm{T}}}{m_{e}c^2} \int\limits^{\infty}_{\gamma_{e}} \int\limits^{\infty} _{\frac{\epsilon_{\gamma}}{4\gamma_{e}(\epsilon_{\gamma}-\gamma_{e})}} d\epsilon_{\gamma} d\omega 
		\frac{n_{\gamma}(\epsilon_{\gamma})}{\epsilon_{\gamma}^3} \frac{n_{\rm{ph}}(\omega)}{\omega^2} \nonumber \\
		& \times  \Biggl\{ \frac{4\epsilon_{\gamma}^2}{\gamma_{e}(\epsilon_{\gamma}-\gamma_{e})}  
		 \ln \Big [ \frac{4\gamma_{e}\omega(\epsilon_{\gamma}-\gamma_{e})}{\epsilon_{\gamma}} \Big ] -8\epsilon_{\gamma}\omega +  \nonumber \\
		 & +  \frac{2(2\epsilon_{\gamma}\omega-1)\epsilon_{\gamma}^2}{\gamma_{e}(\epsilon_{\gamma}-\gamma_{e})} - \left ( 1- \frac{1}{\epsilon_{\gamma}\omega} \right ) \frac{\epsilon_{\gamma}^4}{\gamma_{e}^2(\epsilon_{\gamma}-\gamma_{e})^2} \Biggr\} .
	\end{align}

\noindent Here $\gamma_{e}=E_{e}/m_{e}c^2$ is the Lorentz factor of the electron, $\epsilon_{\gamma}=E_{\gamma}/m_{e}c^2$, and $\omega=\epsilon/m_{e}c^2$ are the dimensionless photon energies. 

Another important source of electron/positron pairs when protons are present is the Bethe-Heitler process. To estimate this contribution, we use the treatment described by \citet{chodorowski1992}.

\subsection{Radiative losses}

Loss terms in our equations include synchrotron radiation, inverse Compton (IC) scattering, and relativistic bremsstrahlung for electrons and muons. We also consider photon production by pair annihilation. For protons, the relevant mechanisms are synchrotron radiation, photomeson production, and hadronic inelastic collisions.

A complete discussion of cooling times due to synchrotron radiation, IC scattering, and hadronic interactions can be found, for example, in \citet{vilaaha} and \citet{flor01}. The injection of secondary particles, such as pions and muons, is also discussed in these works.

The number of pairs is modified by both creation and annihilation. A useful approximation to the annihilation rate is given by \citet{coppi1990},

	\begin{align}
				t^{-1}_{e^{\pm}}(E_{\pm}) = \frac{3}{8} \frac{\sigma_{\rm{T}}c(m_{e}c^2)^2}{E_{\pm}} & \int^{E_{\mp}^2}_{E_{\mp}^1} dE_{\mp} \frac{N_{\mp}(E_{\mp})}{E_{\mp}} \nonumber\\
				& \times \left[ \ln \Big( \frac{4E_{+}E_{-}}{(m_{e}c^2)^2} \Big) -2 \right] ,
		\end{align}		

\noindent where $N_{\mp}(E_{\mp})$ represents the electron/positron distribution (in units of erg$^{-1}$ cm$^{-3}$).

\begin{figure*}
\centering
\subfigure[Electron losses.]{\label{fig:perdidas:a}\includegraphics[width=0.45\textwidth,keepaspectratio]{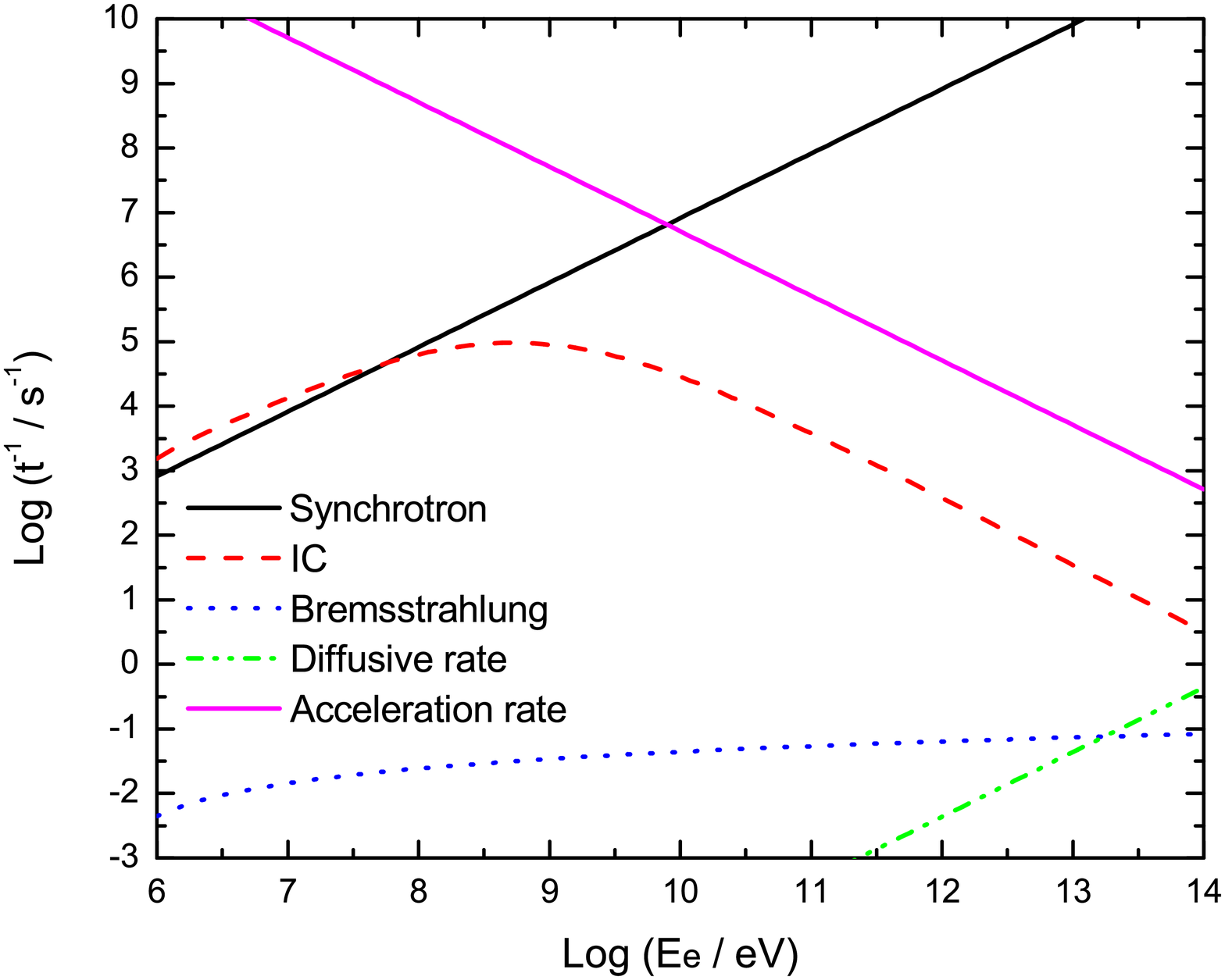}} \hspace{20pt} 
\subfigure[Proton losses.]{\label{fig:perdidas:b}\includegraphics[width=0.45\textwidth,keepaspectratio]{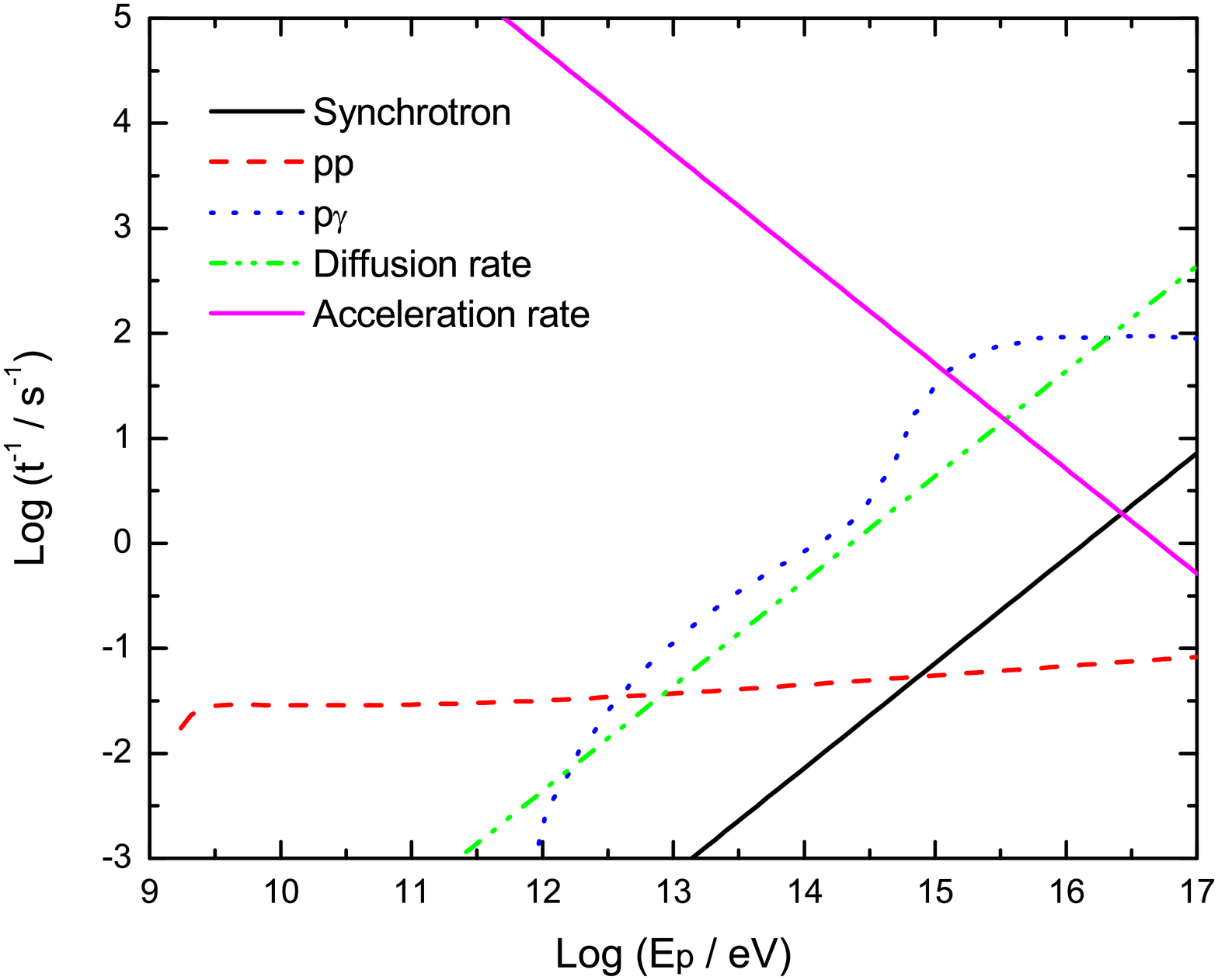}} \hfill \\ 
\subfigure[Pion losses.]{\label{fig:perdidas:c}\includegraphics[width=0.45\textwidth, keepaspectratio]{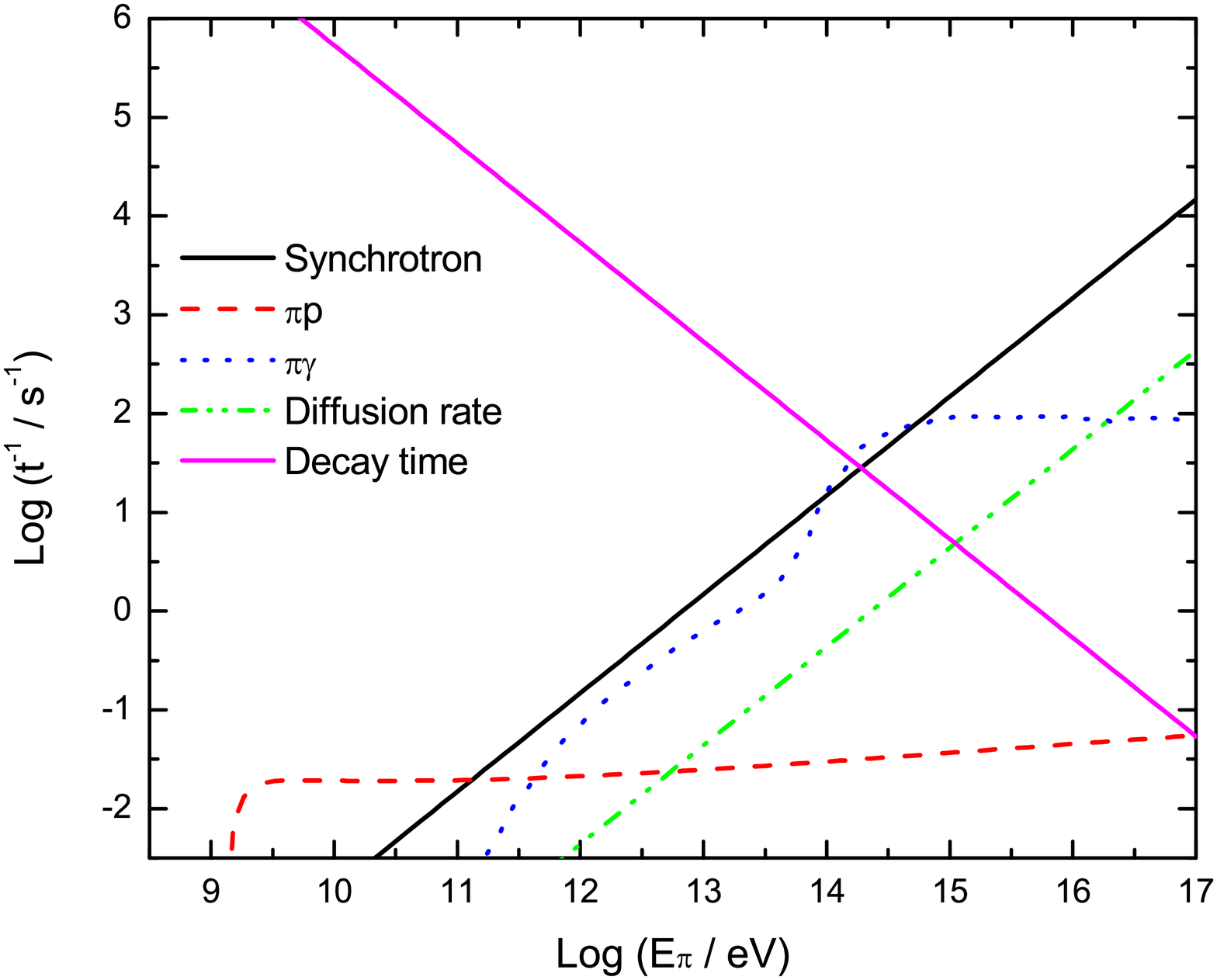}}  \hspace{20pt}
\subfigure[Muon losses.]{\label{fig:perdidas:d}\includegraphics[width=0.45\textwidth, keepaspectratio]{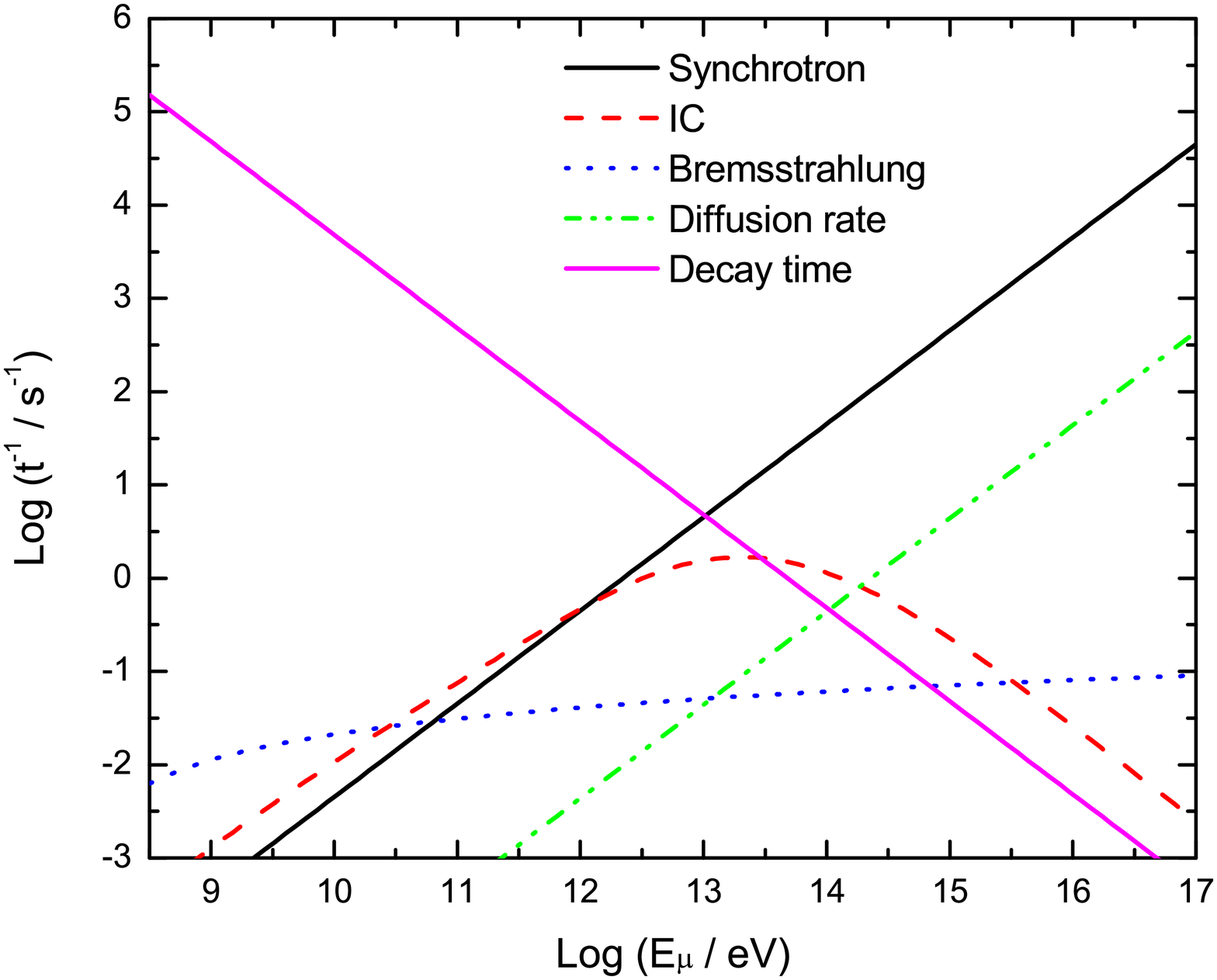}} \hfill
\caption{Relevant radiative losses in a corona characterized by the parameters of Table \ref{table}.}
\label{fig:perdidas}
\end{figure*}

In Fig. \ref{fig:perdidas}, we show the cooling rates together with the diffusion, decay, and acceleration rates of all particle species. The value adopted for $E_{\rm{min}}$ is twice the rest-frame energy of each type of particle. The maximum energy for electrons and protons can be inferred using a balance between the acceleration rate, given by Eq. \ref{eq:accrate}, and the cooling rate. This yields $E^{e}_{\rm{max}} \sim 10$ GeV for electrons, and $E^{p}_{\rm{max}} \sim 10^{3}$ TeV for protons. Particles of such energies satisfy the Hillas criterion, and can be contained within the source.

The cooling rates for the IC scattering and photomeson production shown in the figures are the result of the interactions of particles with two target photon fields: the X-ray emission of the corona and the emission from the disk. 

Cygnus X-1 is a binary system with a massive star that produces an intense radiation field. This photon field could be considered as an additional target for IC scattering since it dominates the bolometric luminosity of the source. However, its effect on particle energy losses is negligible. This can be shown by a simple analysis: in the Thompson regime for IC scattering, the cooling rate ($t^{-1}$) is proportional to the energy density of the target photon field, defined by

\begin{equation}
	 u_{\rm{ph}}= \int{E_{\rm{ph}} n_{\rm{ph}} dE_{\rm{ph}}}.
\end{equation}

\noindent This magnitude for both the accretion disk and the corona X-ray photon fields is on the order of magnitude $\sim 10^{10}$ erg cm$^{-3}$, whereas for the stellar field at the location of the corona it is $\sim 10^{3}$ erg cm$^{-3}$. The photon field provided by the star is then an ineffective seed target for IC processes compared with the local X-ray field.

As can be noted from the figures, no unique mechanism clearly dominates the energy losses for a given particle species. For electrons and muons, there are two relevant radiative processes, IC scattering and synchrotron radiation, whereas for protons and charged pions hadronic interactions are also important. Since IC scattering and photomeson production are processes that depend on the radiation field and, at the same time, the photon field is affected by all the interactions of particles with the different fields, the particle cooling times and distributions are strongly coupled with the transport of photons. This point implies that it is essential to improve the model presented in \citet{flor01}, solving the system of coupled kinetic equations for all type of particles.

\subsection{Set of coupled equations}

We determine the relativistic particle and photon distributions solving the set of coupled transport equations in the steady state and assuming spatial homogeneity and isotropy. The set of kinetic equations are

\begin{enumerate}[a)]

\item{\textsl{Transport of electron/positron pairs and protons:}}

\begin{equation}\label{eq:electron}
		\frac{\partial }{\partial E} \left( b_{i}(E) N_{i}(E) \right)+ \frac{N_{i}(E)}{t_{\rm{esc}}} = Q_{i}(E),
	\end{equation} 
	
\noindent where $i=e^+$, $e^-$, $p$.
	
\item{\textsl{Transport of charged pions and muons:}}

\begin{equation}\label{eq:pion}
		\frac{\partial }{\partial E} \left( b_{i}(E)  N_{i}(E) \right)+ \frac{N_{i}(E)}{t_{\rm{esc}}} + \frac{N_{i}(E)}{t^{i}_{\rm{dec}}} = Q_{i}(E),
	\end{equation} 

\noindent where $i=\pi^+$, $\pi^-$, $\mu^+$, $\mu^-$.

\item{\textsl{Transport of photons:}}

\begin{equation}\label{eq:photon}
	\begin{aligned}
		\frac{N_{\gamma}(E_{\gamma})}{t_{\rm{esc}}^{^{\gamma}}} = Q_{\gamma}(E_{\gamma}) &+ Q_{e^{\pm} \rightarrow \gamma}(N_{e^{\pm}},E_{\gamma}) \\
		&- Q_{\gamma\gamma\rightarrow e^{\pm}}(N_{\gamma},E_{\gamma}) .
		\end{aligned}
	\end{equation}

\end{enumerate}

\noindent Here, $N_{i}(E)$ represents the steady state of each particle distribution (in units of erg$^{-1}$ cm$^{-3}$), $b(E)$ includes all radiative losses for a given type of particle, $t_{\rm{esc}}$ is the timescale over which relativistic particles escape from the system, $t^{i}_{\rm{dec}}$ is the mean decay time for transient particles (pions and muons), and $Q_{i}(E)$ is the injection function. 

In Eq. (\ref{eq:photon}), the term $Q_{\gamma}(E_{\gamma})$ represents photon injection due to several radiative processes

\begin{equation}
Q_{\gamma}(E_{\gamma}) = Q_{\rm{synchr}}(E_{\gamma}) +Q_{\rm{IC}}(E_{\gamma}) +Q_{\pi^0 \rightarrow \gamma\gamma}(E_{\gamma}) .
\end{equation}

\noindent where $Q_{\rm{synchr}}(E_{\gamma})$, $Q_{\rm{IC}}(E_{\gamma})$, and $Q_{\pi^0 \rightarrow \gamma\gamma}(E_{\gamma})$ give the contribution from synchrotron radiation, IC scattering, and neutral pion decay to photon injection, respectively.

The process of pair annihilation is another source of photons. The corresponding annihilation line emissivity can be computed as \citep{svensson1982,boettcher1996}

\begin{align}
Q_{e^{\pm}}(N_{\gamma},E_{\gamma}) = \frac{1}{m_{e}c^2} \int \int & dE_{e^{+}}  dE_{e^{-}} R_{e^{\pm}}(E_{e^{-}},E_{e^{+}},E_{\gamma}) \nonumber\\
 & \times N_{e^{+}}(E_{e^{+}}) N_{e^{-}}(E_{e^{-}})  ,
\end{align}

\noindent where

\begin{equation}
\begin{aligned}
R_{e^{\pm}} & =  \frac{3}{8} \frac{\sigma_{\rm{T}}c (m_{e}c^2)^5}{E_{e^{+}}^2E_{e^{-}}^2} \times\\
& \times \left[ \frac{(\gamma^{\rm{U}}_{\rm{CM}})}{|E_{\gamma}-E_{e^{+}}|+2m_{e}c^2/\pi} + \frac{(\gamma^{\rm{U}}_{\rm{CM}})}{|E_{\gamma}-E_{e^{-}}|+2m_{e}c^2/\pi} \right],
\end{aligned}
\end{equation}

\begin{equation}
\gamma^{\rm{U}}_{\rm{CM}} = \frac{E_{\gamma}}{m_{e}c^2} \Big( \gamma_{+} + \gamma_{-} -E_{\gamma}/m_{e}c^2 \Big) 
\end{equation}

\noindent for $E_{\gamma} > E_{e^{+}}, E_{e^{-}}$ or $E_{\gamma} < E_{e^{+}}, E_{e^{-}}$, or in any other case:

\begin{equation}
\gamma^{\rm{U}}_{\rm{CM}} = \sqrt{ \frac{1}{2} \Big(1 + \gamma_{+}\gamma_{-} + (\gamma_{-}^2-1)^{1/2}(\gamma_{+}^2-1)^{1/2} \Big) },
\end{equation}

\noindent and $\gamma_{+} = E_{e^{+}}/m_{e}c^2$ and $\gamma_{-} = E_{e^{-}}/m_{e}c^2$. We refer to \citet{romeroparedes} and references therein for formulae on radiative processes.

\subsection{Numerical method}\label{numerical}

We use an Adams-Moulton method (see, e.g., \citealt{press1992}) to solve the differential equations in Eqs. (\ref{eq:electron})-(\ref{eq:pion}). This is an implicit multi-step integration method that can reach higher orders than other numerical algorithms; we use in particular a second order method.

Following the scheme described in \citet{vurm2009}, we define an equally spaced grid on a logarithmic scale for the energy of particles

\begin{equation}
\ln E_{i}  =  \ln E_{\rm{min}} + i \cdot \Delta E ,  i \in [0,i_{\rm{m}}],  \\
\end{equation}
\begin{equation}
\ln E^{\gamma}_{l} =  \ln E^{\gamma}_{\rm{min}} + l \cdot \Delta E^{\gamma} ,  l \in [0,l_{\rm{m}}]. 
\end{equation}

\noindent We then obtain a system of linear algebraic equations of the form

\begin{equation}
\sum_{j=1}^{i_{\rm{m}}} A_{ij} \cdot N_{j} = Q_{i} ,
\end{equation}

\noindent with the boundary condition that $N_{i_{\rm{m}}}=0$, which represents $N(E_{\rm{max}})=0$. The matrix $A_{ij}$ contains the particle losses, whereas particle injection is included in the vector $Q_{i}$

\begin{equation}
Q_{i}=
\left( 
\begin{aligned}
	\frac{1}{2}h_{1}&(Q_{1}+Q_{2}) &  \\
	\frac{1}{2}h_{1}&(Q_{1}+Q_{2})&  \\
	\frac{1}{2}h_{2}&(Q_{2}+Q_{3}) &\\
	&\vdots &\\
	\frac{1}{2}h_{m-1}&(Q_{i_{\rm{m}-1}}+Q_{\rm{m}})& \\
	&0 &
\end{aligned} 
\right),
\end{equation}

\noindent where $h_{j} = E_{j+1}-E_{j}$ is the energy step.

	We first solve the transport equations and obtain the particle distributions. These are used to estimate to first order the non-thermal luminosity. Once we know the non-thermal photon injection, we solve Eq. \ref{eq:photon}. An important property of the photon transport equation (\ref{eq:photon}) is its nonlinearity. This is because the cross-section of photopair production ($Q_{\gamma\gamma\rightarrow e^{\pm}}(N_{\gamma},E_{\gamma})$) depends explicitly on the photon distribution. We use the approximation discussed in \citet{poutanen2009}, which consists in taking the photon distribution from a previous step, $j$, to obtain the current injection (step $j+1$) of electron/positron pairs. Since we firstly consider a steady state, the way to solve the photon transport equation is then reduced to a simple iterative scheme given by

\begin{equation}
	\begin{aligned}
		N^{j+1}_{\gamma}(E_{\gamma}) = t_{\rm{esc}}^{^{\gamma}} \Big( Q^{j+1}_{\gamma}(E_{\gamma}) &+ Q^{j+1}_{e^{\pm} \rightarrow \gamma}(N^{j}_{e^{\pm}},E_{\gamma}) \\
		&- Q^{j+1}_{\gamma\gamma\rightarrow e^{\pm}}(N^{j}_{\gamma},E_{\gamma}) \Big) .
		\end{aligned}
	\end{equation}

	The updated photon distribution is then added to the background photon fields (corona power-law plus emission of the disk)\footnote{As we have mentioned, at the location of the corona the stellar photon field can be considered as negligible.} to compute the IC scattering and hadronic interactions.  We calculate the radiative losses and injection of particles where appropriate. The transport equations of massive particles are then solved, and the new distributions are used to compute the luminosity to second order. The process is repeated until all particle distributions converge to a stationary value.

\subsection{Photon absorption in the stellar radiation field}

The binary system Cygnus X-1 is composed of a massive star and a compact object. The massive star is an O9.7 Iab star of $\sim 20 M_{\odot}$ \citep{orosz2011}. The orbit of the system is circular, with a period of 5.6 days and an inclination of between $25^{\circ}$ and $30^{\circ}$ \citep{orosz2011}.

The star produces an intense radiation field that can absorb gamma rays by pair creation within the binary system. The photon field of the star is anisotropic, because its intensity depends on the position of the black hole in its orbit (but outside the corona). The gamma ray absorption in X-ray binaries with a massive companion star has been studied, for example, by \citet{herterich1974}, \citet{carraminana1992}, \citet{bednarek1993,bednarek2000}, \citet{boettcher2005}, \citet{dubus2006}, \citet{zdziarski2009}, and \citet{delValle2010}. To estimate the observable spectrum and compare with the available data, it is necessary to include an appropriate treatment of the absorption in the stellar field. We then use the approach described in \citet{delValle2010}, where the case of Cygnus X-1 is considered.

The star has a radius $R_{*}=1.5 \times 10^{12}$ cm, and we assume a blackbody radiation density of temperature $T_{*}=3 \times 10^{4}$ K. The orbital radius is $r_{\rm{orb}} = 3.4 \times 10^{12}$ cm. Table \ref{estrella} lists the values of the companion star and orbit parameters \citep{orosz2011}.

\begin{table}
    \caption{Orbital and stellar parameters.}
    \label{estrella}
    \centering
\begin{tabular}{ll}
\hline\hline
Parameter & Value\\
\hline\noalign{\smallskip}
$M_{\star}$:   Star mass [$M_{\odot}$]										& $20$  										\\
$R_{\star}$:   Star radius [$R_{\odot}$]  		  						& $17$          						\\
$T_{\star}$:   Star temperature [K] 			   	  					&	$3.0 \times 10^4$						\\
$P_{\rm{orb}}$: Orbital period [days] 											&	$5.6$     								\\
$a$:           Semi-major axis [cm] 											&	$2.3 \times 10^{12}$    	\\
$i$:           inclination angle [$^{\circ}$]				      & $27$   										\\

\hline\\
\end{tabular}	
  
\end{table}

\begin{figure}
  \resizebox{\hsize}{!}{\includegraphics[angle = 270]{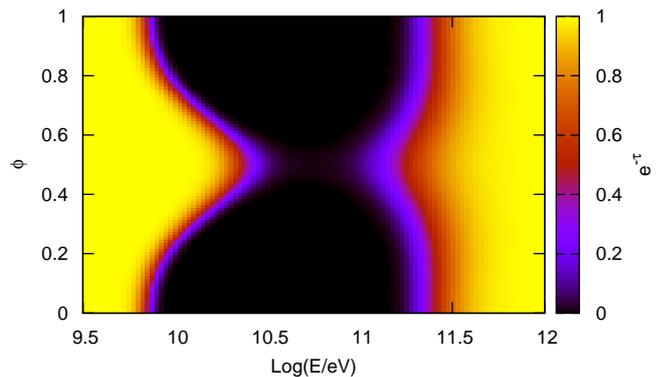}}
\caption{Map of the absorption produced by the anisotropic stellar photon field.}
\label{fig:mapa}
\end{figure}

\begin{figure}
  \resizebox{\hsize}{!}{\includegraphics{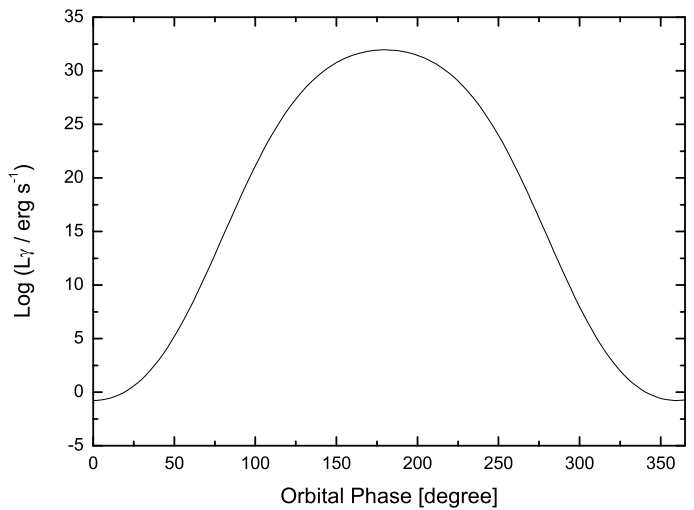}}
\caption{Modulation of gamma ray emission in the steady state at $E \sim 50$ GeV by the anisotropic photon field of the companion star. }
\label{fig:modulation}
\end{figure}

In Fig. \ref{fig:mapa}, we show the opacity map produced by the photon absorption in the stellar field, in the relevant energy range and along the complete orbit ($\phi$ is the orbital phase in units of $2\pi$; we note that $\phi=0=1$ corresponds to the compact object at opposition, i.e. superior conjunction). As can be seen in the figure, except for close to the inferior conjunction (compact object in front of the star, $\phi = 0.5$), the emission is completely suppressed by the stellar photon field at energies in the range $10$ GeV $< E < 120$ GeV. To better illustrate this effect, Fig. \ref{fig:modulation} shows the modulation with the orbital phase of gamma ray emission of the corona in a steady state at a given energy. The stellar photon field is almost transparent when the compact object passes in front of the star ($\phi = 0.5$), but almost opaque when the compact object passes behind ($\phi = 0$). These results agree with those obtained by \citet{delValle2010}.


\subsection{Spectral energy distribution}

\begin{figure}
  \resizebox{\hsize}{!}{\includegraphics{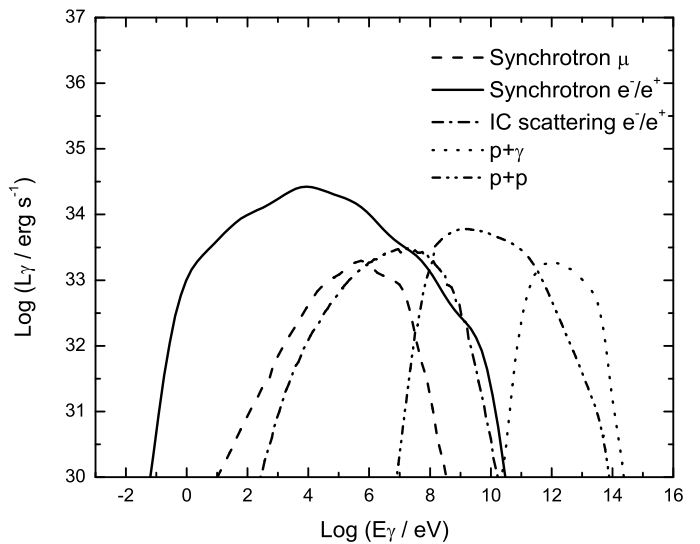}}
\caption{Main non-thermal contribution to the luminosity, without considering internal absorption. }
\label{fig:mainSEDs}
\end{figure}

Figure \ref{fig:mainSEDs} shows the main non-thermal contributions to the total luminosity. The internal absorption is not included in this plot. The synchrotron radiation of electron/positron pairs dominates the spectrum for $E_{\gamma} < 100$ MeV. We note that given the small size of the corona, the synchrotron radiation below $E < 1$ eV is self-absorbed. All radio emission of the source comes from the jet \citep{stirling2001}.

 The relativistic protons injected in the corona collide with thermal protons through different channels ($\zeta_{1}$ and $\zeta_{2}$ are the multiplicities)

\begin{equation}
				p+p \rightarrow p+p+\zeta_{1}\pi^0 +\zeta_{2}(\pi^++\pi^-)\textrm{,}
			\end{equation}
	
			\begin{equation}
				p+p \rightarrow p+n+\pi^++\zeta_{1}\pi^0 +\zeta_{2}(\pi^++\pi^-)\textrm{,}
			\end{equation}
			
			\begin{equation}
				p+p \rightarrow n+n+2\pi^++\zeta_{1}\pi^0+\zeta_{2}(\pi^++\pi^-)\textrm{,}
			\end{equation}
		
\noindent and with the photon field
			
			\begin{equation} \label{eq:pgamma1}
				p+\gamma \rightarrow p+\zeta_{1}\pi^0+\zeta_{2}(\pi^++\pi^-) \textrm{,}
			\end{equation}
			
			\begin{equation}\label{eq:pgamma2}
				p+\gamma \rightarrow n+\pi^+ +\zeta_{1}\pi^0+\zeta_{2}(\pi^++\pi^-) \textrm{.}
			\end{equation}
			
\noindent The main electromagnetic result of these interactions is the neutral pion decay

\begin{equation}
		\pi^{0} \rightarrow \gamma + \gamma.
	\end{equation}
	
\noindent This is the most relevant source of photons in the high-energy gamma-ray band (see Fig. \ref{fig:mainSEDs}).

One assumption of our model is the equipartition of energy between the magnetic energy density and the photon energy density of the corona. It is then expected that the contribution of synchrotron radiation and IC scattering to the total luminosity be comparable. In the analysis of Fig. \ref{fig:perdidas}, we showed that IC scattering and synchrotron radiation are the radiative processes dominating the energy losses for electrons at low energies. At higher energies, the Klein-Nishina effect becomes important, the IC cross-section decreases, and this leaves synchrotron as the main mechanism causing electron energy loss. The Klein-Nishina effect is also responsible for the diminution of the IC radiation with respect to the Thompson regime; the gamma-ray flux is proportional to the number of interactions, hence when the IC cross-section decreases, so does the gamma ray emission. This explains why in Fig. \ref{fig:mainSEDs} synchrotron radiation dominates the luminosity of the source at low energies. Nevertheless, the IC emission of electron/positron pairs is comparable to the synchrotron radiation at $E \sim 10^{7-8}$ eV.
		
\begin{figure*}
\centering
\hfill
\subfigure[$a=100$]{\label{fig:dis:a}\includegraphics[width=0.45\textwidth, keepaspectratio]{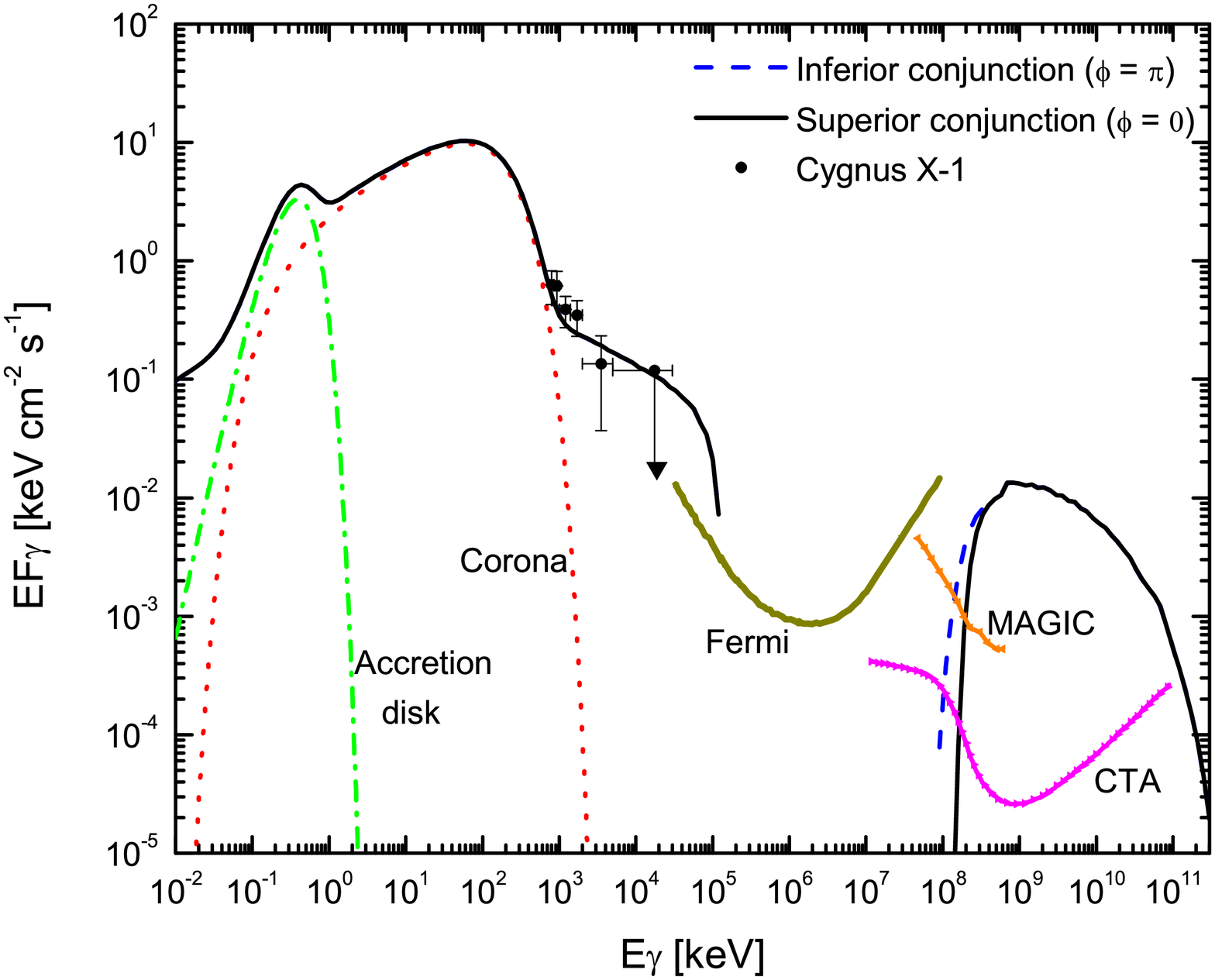}} \hspace{20pt}
\subfigure[$a=1$]{\label{fig:dis:b}\includegraphics[width=0.45\textwidth, keepaspectratio]{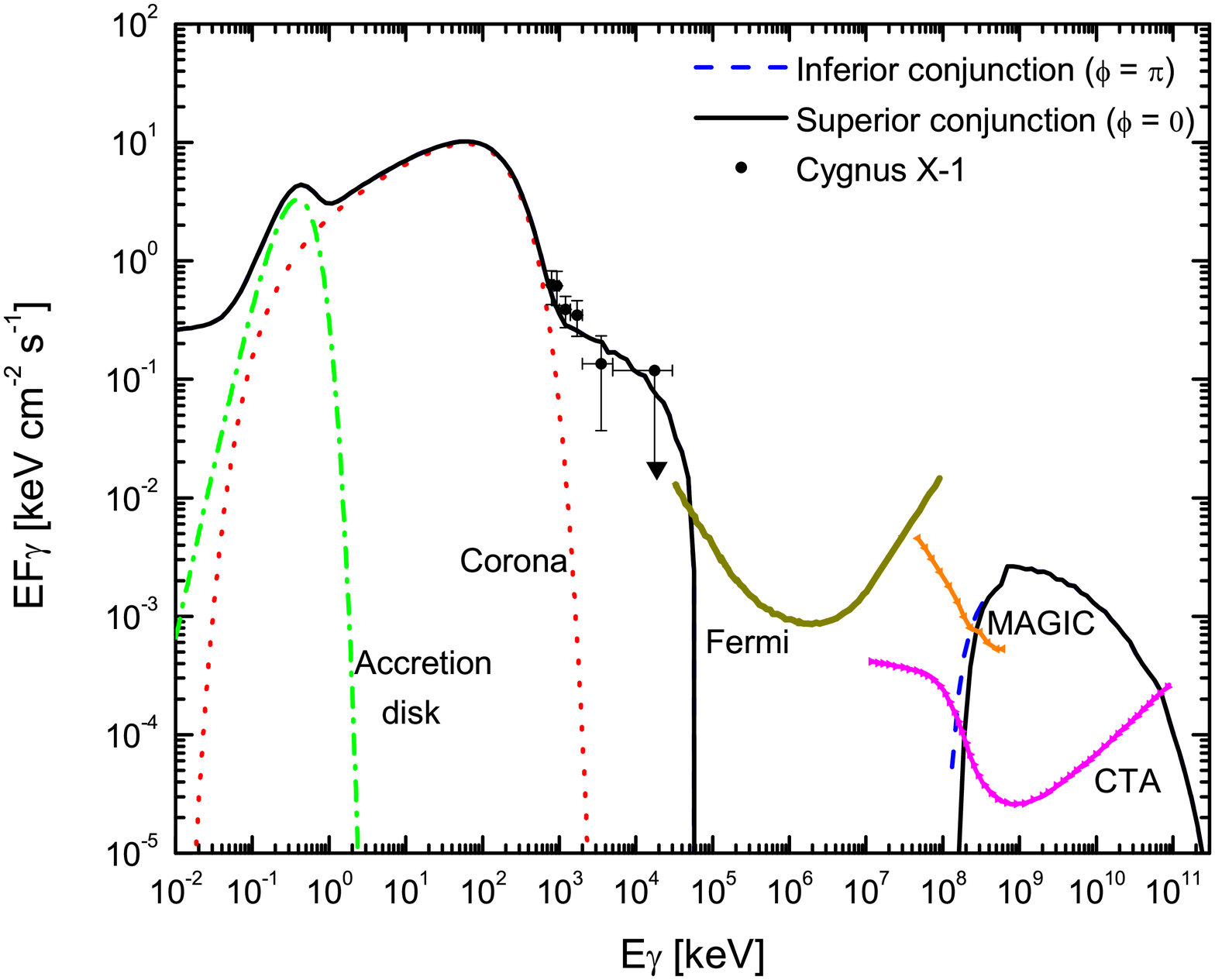}} \hfill \\
\caption{Final flux in a corona + disk characterized by the parameters of Table \ref{table}.  We include the $5\sigma$ sensitivities for different instruments (50 hours of direct exposure for MAGIC and CTA and 1 yr survey mode for Fermi).}
\label{fig:luminosidad}
\end{figure*}

In Fig. \ref{fig:luminosidad}, we show the total photon flux produced in the corona in two different orbital phases, superior and inferior conjunction. These positions correspond to the maximum and minimum absorption in the stellar photon field, respectively. We also show the SEDs obtained for two different values of the parameter $a$. We compare our results with observations of Cygnus X-1 made by COMPTEL \citep{McConnell}, obtaining good agreement. The best-fit model is obtained with $q_{\rm{rel}}=0.02$ for a corona dominated by protons ($a=100$) and $q_{\rm{rel}}=0.03$ for $a=1$; both values are significantly lower that the maximum energy available for particle acceleration. 

\citet{jourdain2011} reported observations of Cygnus X-1 made with the SPI instrument onboard the INTEGRAL satellite. Despite no significant emission above 1 MeV being detected by INTEGRAL, the upper limits are consistent with the non-thermal tail observed at several MeV by COMPTEL, and in close agreement with our model.

\begin{figure}
  \resizebox{\hsize}{!}{\includegraphics{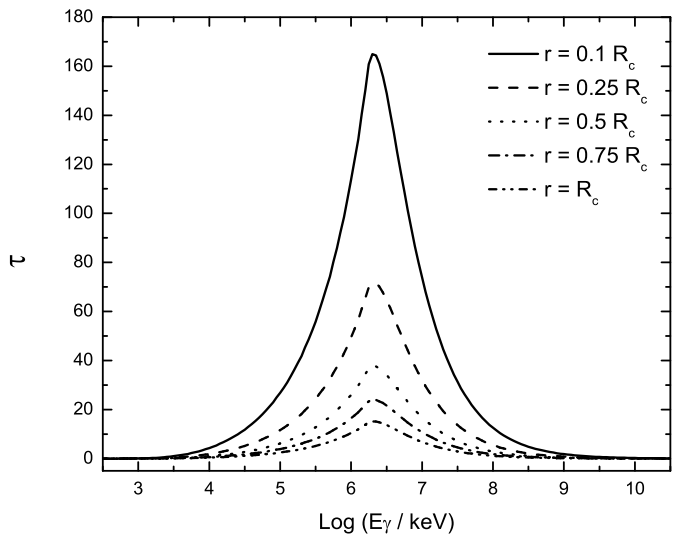}}
\caption{Internal absorption due to photopair production in the soft photon field of the corona and the accretion disk.}
\label{fig:opacity}
\end{figure}

The gap observed in the energy range $10^5 < E < 10^8$ keV is produced by the internal absorption in the corona and accretion disk fields. To quantify this effect, we show in Fig. \ref{fig:opacity} the opacity as a function of the photon energy at different depths inside the corona. Given the high values of the opacity, the emission is completely suppressed. This result is in accordance with the non-detection of Cygnus X-1 in a steady state by Fermi. As pointed out in \citet{flor01}, all emission detected in this energy range should be produced in the jet (e.g., \citealt{Bosch-Ramon2008}).

The absorption in the stellar field partially suppresses the high-energy bump at $E \sim 10^{10-11}$ eV, which makes it difficult to detect this source using either the MAGIC or VERITAS Cherenkov telescopes. The high-energy emission may be detectable by future instruments with higher sensitivity and wider energy ranges, such as the Cherenkov Telescope Array (CTA). 

\section{Transient episode or flare}\label{flare}

It is well-known that X-ray binaries undergo transient radiative episodes. In 1999, the BATSE instrument detected an increase in the luminosity above $50$ keV in Cygnus X-1, of an order of magnitude \citep{stern2001}. Moreover, between 1995 and 2003, seven outbursts have been reported at the location of this source with a significance of 3$\sigma$ or more \citep{mazets1996,romero2002,golenetskii2003}. The luminosities above 15 keV of the outbursts were in the range $1 - 2 \times 10^{38}$ erg s$^{-1}$, which are much higher than the typical thermal luminosity in the hard state.

More recently, the claimed $4.2 \sigma$ detection of Cygnus X-1 during a flaring state by \citet{albert2007} and the suggested detection by the \textsl{AGILE} satellite \citep{tavani2009} constitute the first presumed evidence of very high-energy gamma ray emission produced around a Galactic black hole.

Another example is the X-ray binary Cygnus X-3, from which four gamma-ray flares were detected by \textsl{AGILE} satellite \citep{tavani2009}. Variable emission was indeed detected by \textsl{Fermi} \citep{abdo2009}. This emission, however, is likely produced in the jet \citep{bednarek2010,araudo2010,cerutti2011}.

With the aim of studying transient events in the corona, we apply our model to the case of a non-thermal flare, i.e. a flare produced by changes in the total power injected into relativistic particles ($q_{\rm{rel}}$). The dynamical element during these type of outbursts is the magnetic field; a sudden injection might be the result of both fast and large-scale reconnection events. This assumption is supported by observations of solar flares that suggest that magnetic reconnection can trigger diffusive acceleration without the requirement of strong shock formation \citep{tsuneta1998,lin2008,kowal2011}. The overall thermal luminosity can remain constant during these episodes.

\subsection{Particle injection}

Although the light curves of Galactic black holes during outbursts or transient episodes can vary from source to source, there are some common features. The rise time tends to be much shorter than the decay time \citep{grove1998}, thus the light curves are usually called FRED (Fast Rise and Exponential Decay). A simple analytic expression that can represent this behavior is given by \citep{reynoso2010}

\begin{align}
Q(E,t) = & Q_{0} E^{-\alpha} e^{-E/E_{\rm{max}}} (1-e^{t/\tau_{\rm{rise}}} ) \nonumber \\
& \times \left[ \frac{\pi}{2}- \arctan \Big( \frac{t-\tau_{\rm{plat}}}{\tau_{\rm{dec}}} \Big) \right],
\end{align}

\noindent where $\tau_{\rm{rise}}$, $\tau_{\rm{dec}}$, and $\tau_{\rm{plat}}$ are the rise, decay, and plateau duration, respectively. We adopt $\tau_{\rm{rise}} = 30$ min, $\tau_{\rm{dec}} = 1$ h, and $\tau_{\rm{plat}} = 2$ h for a rapid flare. The power-law has the standard index of $\alpha=2.2$. The normalization constant $Q_{0}$ can be obtained from the total power injected into relativistic protons and electrons, $L_{\rm{rel}}=L_{p}+L_{e}$. This power is assumed to be a fraction of the luminosity of the corona $L_{\rm{rel}}=q_{\rm rel} L_{\rm{c}}$. As mentioned in the previous section, in the steady state the best fit to the observations is obtained with $q_{\rm rel}=0.02$ for $a=100$. During the flare, the number of relativistic particles increases. In our model, the power injected into the flare doubles that injected in the steady state, but larger flares are quite possible, as observed in the Sun \citep{lin2008}. It is assumed that the thermal corona remains unaffected during the event.

\subsection{Spectral energy distribution}

In Galactic black hole coronae, cooling timescales are significantly shorter than the flare timescales, which are typically of hours or even days \citep{malzac2000}. Consequently, the transport equation could be equally solved assuming a steady state, and considering changes in the luminosity as the flare evolves. Since one of the aims of this work had been to develop a code that can be applied to different environments, we include the time-dependent term in the kinetic equations. The transport equations then have the following form \citep{ginzburg}

\begin{equation}\label{eq:transporte}
		\frac{\partial N_{i}(E,t)}{\partial t} +  \frac{\partial }{\partial E} \Big( b(E) N_{i}(E,t) \Big)+ \frac{N_{i}(E,t)}{t_{\rm{esc}}}=Q_{i}(E,t) ,
	\end{equation} 

\noindent where as before

\begin{equation}
b(E)= \frac{dE}{dt} \Big | _{\rm{loss}}. 
\end{equation}

\noindent We solve the set of coupled equations using the treatment described in Sec. \ref{numerical}, but now including the time dependence so the computing time significantly increases.

\begin{figure}
  \resizebox{\hsize}{!}{\includegraphics{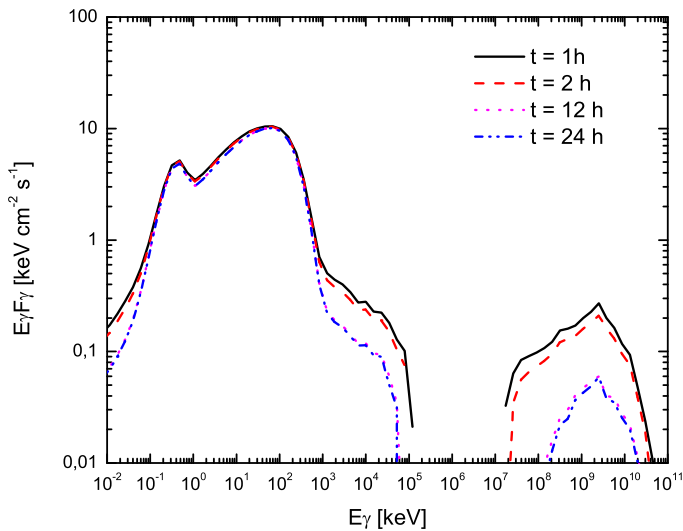}}
\caption{Evolution of the luminosity during a moderate non-thermal flare. We adopt $a=100$. Only the absorption in the corona photon field is considered here, since the transport of photons is solved self-consistently. The absorption in the stellar field, however, is not included in these plots.}
\label{fig:SEDevolution}
\end{figure}

\begin{figure*}
\centering
\subfigure[$\phi_{0} = 0.0$]{\label{fig:tauExterna:a}\includegraphics[width=0.45\textwidth,keepaspectratio]{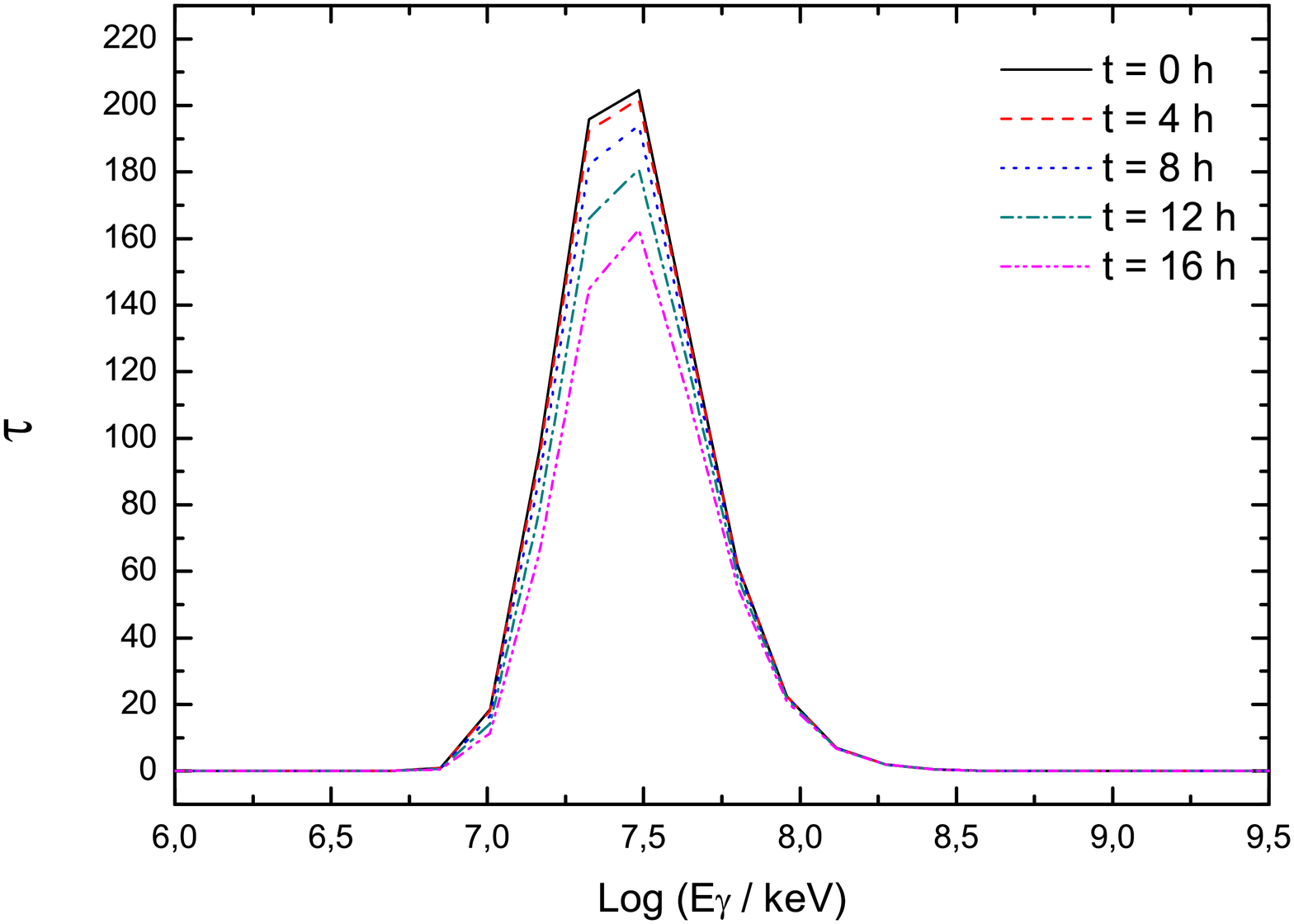}} \hspace{20pt} 
\subfigure[$\phi_{0} = \pi/2$]{\label{fig:tauExterna:b}\includegraphics[width=0.45\textwidth,keepaspectratio]{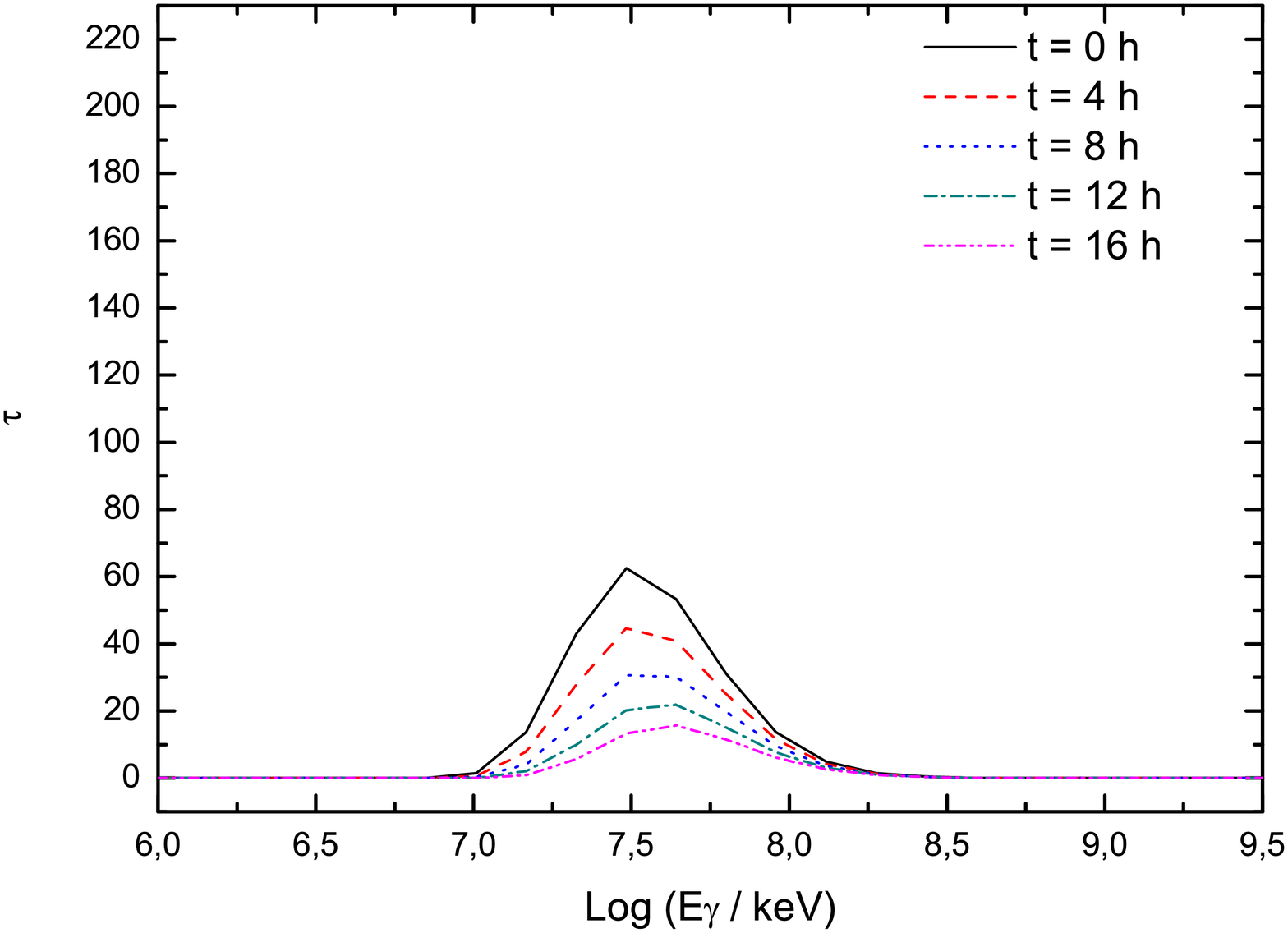}} \hfill \\ 
\subfigure[$\phi_{0} = \pi$]{\label{fig:tauExterna:c}\includegraphics[width=0.45\textwidth, keepaspectratio]{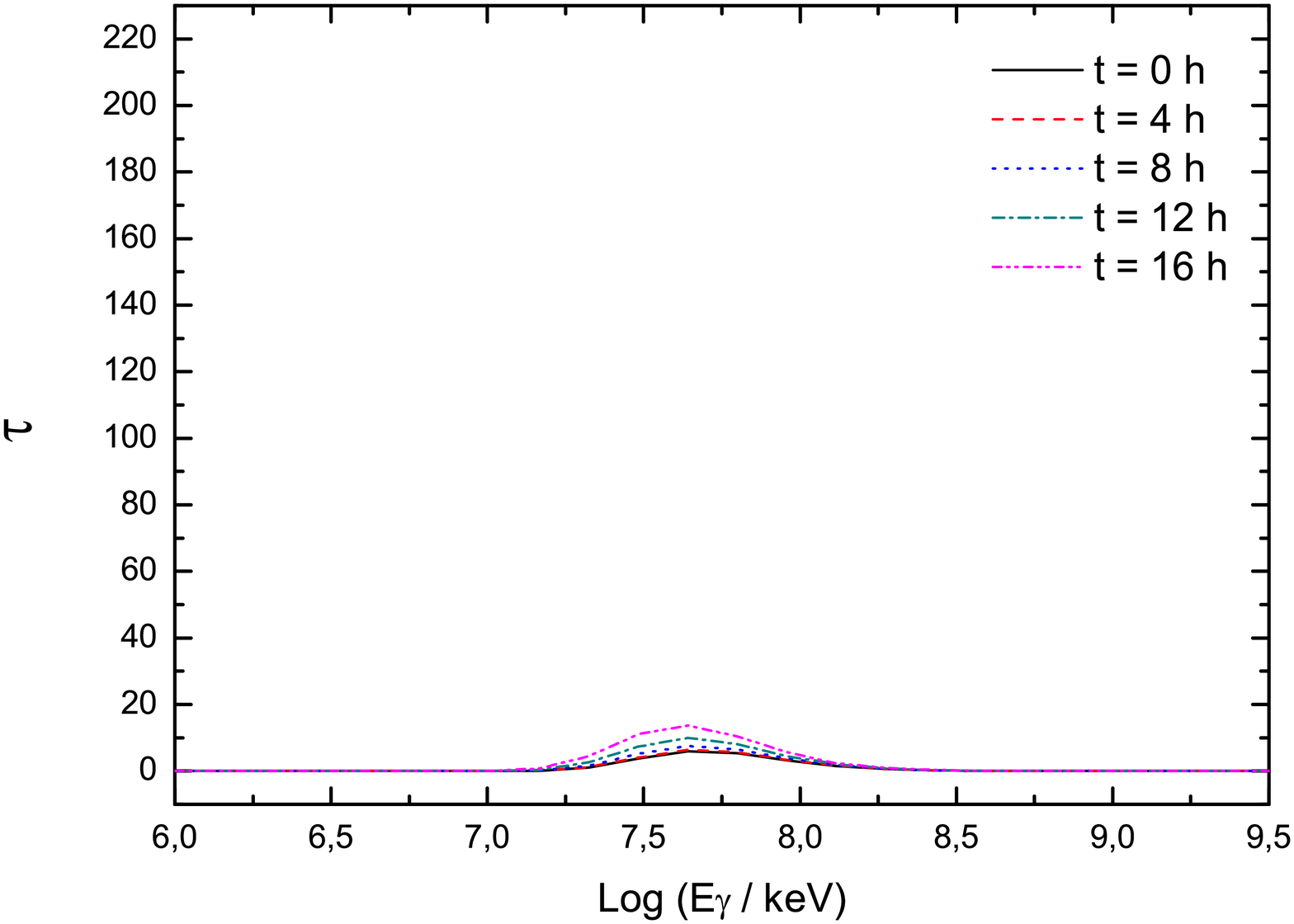}}  \hspace{20pt}
\subfigure[$\phi_{0} = 3\pi/2$]{\label{fig:tauExterna:d}\includegraphics[width=0.45\textwidth, keepaspectratio]{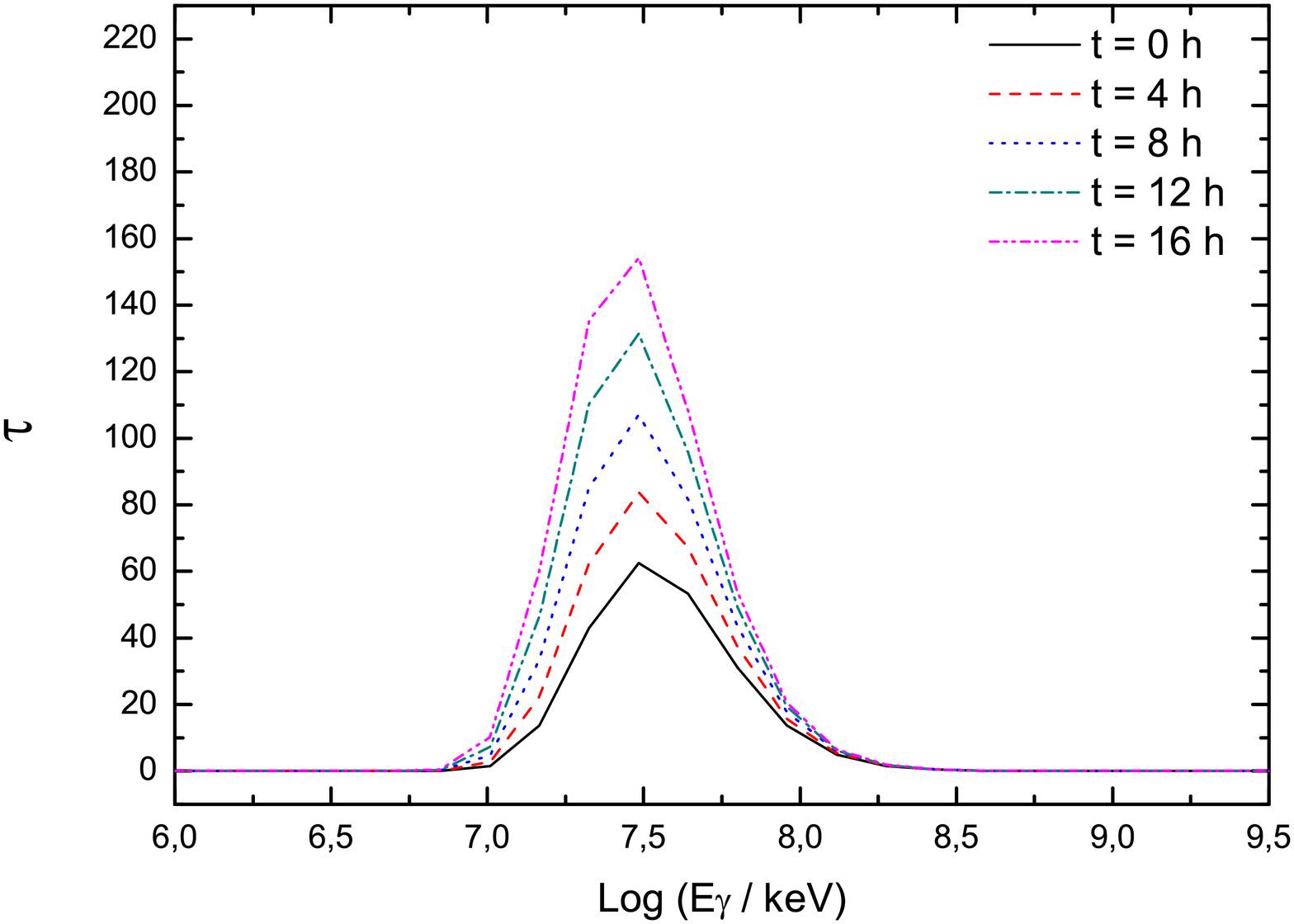}} \hfill
\caption{Changes in the opacity to gamma-ray propagation for flares produced at different orbital phases. All flares are assumed to last ten hours, this is $\sim 7.5$ \% of the orbital period.}
\label{fig:tauExterna}
\end{figure*}

Figure \ref{fig:SEDevolution} shows the evolution of the electromagnetic emission during a day. In this figure, we do not include the absorption in the photon field of the star, since it depends on the orbital phase where the flare is produced. For this purpose, in Fig. \ref{fig:tauExterna} we show the absorption coefficient related to the stellar field. We estimate the opacity for flares occurring at different orbital phases, and we conclude that for some values of $\phi$ the absorption of the star is almost negligible. For example, flares at energies above $10$ GeV can be detected at phases $\phi \sim \pi$, with instruments such as CTA.

\subsection{Neutrino emission}

The electromagnetic flare may be absorbed above $E>10$ MeV. This is not true for a neutrino burst. If a sudden injection of relativistic protons occured, the neutrino flux produced could be detectable by instruments such as IceCube. 

We consider neutrino production by two main channels: charged pion decay

\begin{equation}
		 \pi^{\pm} \rightarrow \mu^{\pm} + \nu_{\mu}(\overline{\nu}_{\mu}) \textrm{,}
		\end{equation}

\noindent and muon decay

	\begin{equation}
		 \mu^{\pm} \rightarrow e^{\pm} + \overline{\nu}_{\mu}(\nu_{\mu}) + \nu_{e}(\overline{\nu}_{e}) .
		\end{equation}
		
\noindent Thus, the total emissivity of neutrinos is \citep{lipari2007,reynoso2009}

\begin{equation}
Q_{\nu}(E,t) = Q_{\pi \rightarrow \nu}(E,t) + Q_{\mu \rightarrow \nu}(E,t) ,
\end{equation}

\noindent where

	\begin{equation}
	\begin{aligned}
		Q_{\pi \rightarrow \nu}(E,t) = \int^{E^{\rm{max}}}_{E} & dE_{\pi} t^{-1}_{\pi,\rm{ dec}}(E_{\pi})N_{\pi}(E_{\pi},t) \times \\
		& \times \frac{\Theta(1-r_{\pi}-x) }{E_{\pi}(1-r_{\pi})},
		\end{aligned}
	\end{equation}
	
\noindent with $x=E/E_{\pi}$, $r_{\pi}=(m_{\mu}/m_{\pi})^2$ and

	\begin{align}
		Q_{\mu \rightarrow \nu}(E,t) &= \sum^4_{i=1} \int^{E^{\rm{max}}}_{E} \frac{dE_{\mu}}{E_{\mu}} t^{-1}_{\mu,\rm{ dec}}(E_{\mu})N_{\mu_{i}}(E_{\mu},t)\\
		 &\times \left[ \frac{5}{3}-3x^2+ \frac{4}{3}x^3 \right]. \nonumber
	\end{align}
	
\noindent In this latter expression, $x=E/E_{\mu}$, $\mu_{\{1,2\}}=\mu_{\rm{L}}^{\{-,+\}}$, and $\mu_{\{3,4\}}=\mu_{\rm{R}}^{\{-,+\}}$. Our calculations take into account pion and muon losses, as in \citet{reynoso2009}.

The differential flux of neutrinos arriving at the Earth can be obtained as

\begin{equation}
\frac{d \Phi_{\nu}}{dE} = \frac{1}{4\pi d^2} \int_{\rm{V}}{d^3r Q_{\nu}(E,t)} .
\end{equation}

\begin{figure*}
  \centering
  \includegraphics[clip,width=0.6\textwidth, keepaspectratio]{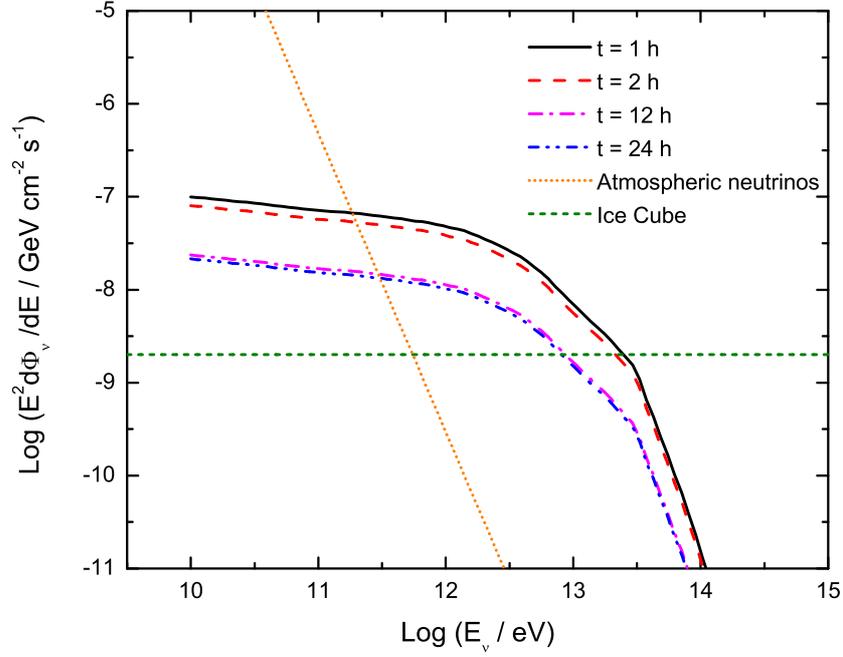}
\caption{Neutrino flux, atmospheric neutrino flux, and IceCube sensitivity. Case with $a=100$.}
\label{fig:neutrinoFlux}
\end{figure*}

\noindent This quantity, weighted by the squared energy, is shown in Fig. \ref{fig:neutrinoFlux}, which also shows the IceCube sensitivity for one year of operation. Assuming that the duty cycle of flares in Galactic black holes is around $10$ \%, the IceCube detector will be able to detect neutrinos from a source at $\sim 1.8$ kpc \citep{reid2011} after ten years of observations.

Nonetheless, a variability search with carefully binned time spans might yield positive results long before (\citealt{vieyroIcecube}, in preparation), avoiding the disadvantage of the smoothing cause by the averaging of the observations.

\section{Discussion}

We have developed a model to deal self-consistently with the non-thermal emission from a magnetized corona. The assumption of equipartition among the magnetic energy density, the bolometric photon energy density of the corona, and the kinetic energy density of the plasma is used to estimate the value of the most relevant parameters. This is supported by the X-ray binaries spending a significant amount of time in the low-hard state, permitting in turn a significant amount of time for field-particle interactions. Under this assumption, the model presented in this work is self-similar, because if a value of one parameter is changed, the other parameters can be re-scaled, producing no significant differences to the results and predictions. 

On the other hand, we found that if we consider a sub-equipartition magnetic field, the radiative output can be very different. The role of the magnetic field in the cooling of the different types of relativistic particles in the corona is very important. If lower values of the magnetic field are considered, the main radiative losses will be produced by the interaction with matter (e.g., relativistic bremsstrahlung), hence the spectra of relativistic particles will be modified. 

The hadronic content of the plasma, given by the parameter $a$, is unknown. We have adopted two values: $a=100$, which represents a proton-dominated corona, and $a=1$, i.e. a corona with equal contributions from both protons and electrons. Both models are capable of reproducing the non-thermal spectrum observed by COMPTEL and INTEGRAL in Cygnus X-1. The main difference is caused by the energy injected into relativistic particles: the lower the hadronic content, the higher the power injected in relativistic particles. Neutrino production increases with $a$.

A significant amount of flux in the range 10 MeV - 1 TeV is not expected because of absorption in the thermal photon field. We instead predict the existence of a bump at very high energies ($E \geq 1$ TeV). This high-energy emission may be detectable in the future from different sources. Since it is of hadronic origin, detections or upper limits can be used to place constraints on the number of relativistic protons in the corona. Orbital modulation can be important owing to the variable absorption \citep{delValle2010}.

For sources where it is difficult to detect an electromagnetic flare, the neutrino production can nevertheless yield detectable events. Our results show that there may be instances in which a neutrino flare could be detected, but the gamma-ray counterpart is not.  

\citet{petropoulou2011,petropoulou2012} studied the temporal signatures of electron and proton injection in a compact, magnetized source. Both photon quenching and synchrotron emission are responsible for the dynamical non-linear behavior of the system. Although these results cannot be directly extrapolated to our corona model, since $pp$ interactions and background thermal photon fields cannot be ignored, future research may identify evidence of some kind of dynamical cycle in sources such as Cygnus X-1.

\section{Conclusions}

We have illustrated that a consistent treatment of the non-thermal emission from a magnetized corona can be implemented by solving the set of coupled differential equations for all particle species.

Our application to Cygnus X-1 provided both a good fit to the observational data and interesting predictions for very high energy and neutrino instruments. 

In the future, we will explore flare episodes taking place in low-mass X-ray binaries. These are attractive objects for the application of our model, because in these systems the companion star has a weak radiation field. Since the absorption will be negligible, both the electromagnetic and the neutrino flares may be detectable.

\section*{Acknowledgments}

We thank Mar\'{\i}a Victoria del Valle for help in several aspects of this work. F.L.V. also thanks Abel Valente and Gabriela Vila for their help in performing the numerical simulations and optimization of the computational implementations. This work was partially supported by the Argentine Agencies CONICET (PIP 0078) and ANPCyT (PICT 2007-00848), as well as by Spanish Ministerio de Ciencia e Innovaci\'on (MICINN) under grant AYA2010-21782-C03-01.

\bibliographystyle{aa}	
\bibliography{myrefs}		

\end{document}